# Mars Thermospheric Polar Warming at Aphelion: Dynamical Processes Studied Using M-GITM

Jia-Zheng Li[1,2] ⬤, Stephen W. Bougher[2] ⬤, Cheng Li[2], and Erdal Yiğit[3,4] ⬤

[1]Macau University of Science and Technology, Macao, China, [2]University of Michigan, Ann Arbor, MI, USA, [3]George Mason University, Fairfax, VA, USA, [4]ITM Lab (675), NASA Goddard Space Flight Center, Greenbelt, MD, USA

**Abstract** Aphelion Thermospheric Polar Warming (TPW), first identified in 2024 by Mars Atmosphere and Volatile EvolutioN (MAVEN) observations, is a dynamical heating phenomenon in the Martian atmosphere that exists in the winter hemisphere near the aphelion solstice. Studying the formation mechanism of aphelion TPW will help us better understand the energy budget of the Martian thermosphere. In this study, we investigate aphelion TPW using the Mars Global Ionosphere Thermosphere global circulation model (M-GITM). The simulation results show that the local dust storms have little impact on the formation of aphelion TPW. The simulated thermospheric temperature difference between the polar region and the low-latitude region is considerably lower than the value observed, which suggests that some important atmospheric processes are not captured by M-GITM. To investigate potential causes, we conduct sensitivity tests on solar insolation, gravity waves, and model horizontal resolution. The sensitivity test on solar insolation shows that the magnitude of aphelion TPW increases with increasing solar insolation. We also find that gravity waves play a critical role in modulating dynamical heating, as their suppression increases the latitudinal temperature difference. Model resolution has minimal impact on polar warming but affects thermospheric structure at low latitudes. These findings highlight the importance of refining the representation of dynamical processes, especially the parameterization of subgrid-scale internal gravity waves in the Martian general circulation model to better capture thermospheric dynamics.

**Plain Language Summary** Thermospheric Polar Warming (TPW) is a warming phenomenon that occurs in the upper atmosphere over the winter polar region. This phenomenon was previously found to occur only during northern winter (solar longitude Ls ≈ 270°, when Mars is closest to the Sun) but not during southern winter (solar longitude Ls ≈ 90°, when Mars is farthest from the Sun). However, MAVEN observations first identified aphelion TPW in 2024. In this study, we investigate the causes of this warming using a Martian global model. Our results show that local dust storms have minimal influence, whereas gravity waves play a critical role. When gravity waves are suppressed in the model, the polar temperature contrast increases, aligning more closely with the observations. These findings suggest that an improved representation of gravity wave processes is essential for accurately capturing Martian thermospheric dynamics.

## 1. Introduction

Since planetary thermospheres are the key region of atmosphere-ionosphere coupling and low-orbit spacecraft drag, the structure of the Mars planetary thermosphere has drawn widespread interest in the planetary and space physics community. The whole atmosphere system of Mars is strongly driven by dust activity, upward propagating internal waves, and space weather (Yiğit, 2023). Thermospheric Polar Warming (TPW), which occurs in the winter hemisphere near the solstices, is one of the important phenomena that exist in the Martian thermosphere. TPW was first observed in the northern winter hemisphere during near perihelion conditions using measurements from the Mars Odyssey orbiter (Bougher et al., 2006). However, corresponding TPW near the aphelion season could not be identified by this study. Based on the solar occultation measurements made by the Extreme Ultraviolet Monitor (EUVM) onboard the Mars Atmosphere and Volatile EvolutioN (MAVEN) spacecraft, Thiemann et al. (2024) first identified TPW during aphelion and showed that this TPW was present at dawn but not at dusk. The finding that TPW can also occur under conditions of minimum solar insolation is intriguing. Along the morning terminator, the observed thermospheric temperature difference is more than 40 K between the south polar region (around 65°S) and the lower-latitude region (around 40°S), showing that there is a significant amount of heating at the dawn side of the south polar region. Thiemann et al. (2024) also indicated that the intensity of TPW exhibits annual variations, which may be caused by the variation of a northern-hemisphere local dust storm. Several studies have successfully simulated the existence of perihelion TPW using a general circulation model (GCM) and indicated that several factors may influence the properties of TPW. Bell, Bougher, and Murphy (2007) showed that the lower atmospheric solar tide can shift TPW poleward by 10°–20° and González-Galindo et al. (2009) showed that the in situ tides in the thermosphere are crucial for the existence of TPW. Bell et al. (2007) also showed that the increasing dust optical depth in the summer hemisphere may enhance the magnitude of TPW. Medvedev and Yiğit (2012) and Medvedev et al. (2013) demonstrated that internal gravity waves in the Martian atmosphere modulate the thermospheric heating and enhance TPW. Gravity waves are routinely observed throughout the Martian whole atmosphere system during all seasons (Heavens et al., 2022; Jesch et al., 2019; Starichenko et al., 2024; Yiğit et al., 2021; Yiğit, Medvedev, & Hartogh, 2021). As ubiquitous features of planetary atmospheres (Yiğit & Medvedev, 2019), they are a major driver of atmospheric circulation and therefore play an important role in TPW.

Due to the absence of the solar insolation at high winter latitudes, TPW is dominated by the dynamical heating in the thermosphere. Dynamical heating effects have successfully explained several phenomena in planetary thermospheres, for example, the cooling in the polar thermosphere on Earth (Crowley et al., 1995; Schoendorf et al., 1996), the warm temperature in the night-side lower thermosphere during the long Venusian night (Brecht et al., 2011), and the occurrence of regions with significant thermospheric temperature and density enhancements near the morning and evening terminators on Mars (Forbes & Moudden, 2009; Pilinski et al., 2018). Pilinski et al. (2023) reviewed the dynamical heating effect in the Martian thermosphere by comparing the statistical results from MAVEN observations and the simulation results from the Mars Global Ionosphere Thermosphere Model (GITM) (M-GITM). They showed that the dynamical heating effect is most prominent in the regions of converging winds near the morning and evening terminators (Pilinski et al., 2023), which is in good agreement with previous studies. Since the TPW is one of the key phenomena in the Martian thermosphere, its mechanisms are important for us to understand in addressing the overall role of thermosphere dynamics. To examine the dynamical heating effect during aphelion TPW and to explore the mechanisms of aphelion TPW, we perform systematic global-scale simulations with M-GITM (Bougher et al., 2015), incorporating a state-of-the-art whole atmosphere gravity wave parameterization (Yiğit et al., 2008).

The structure of the paper is as follows. Section 2 summarizes the recent observations of the Mars Year (MY) 34 aphelion TPW. Section 3 describes M-GITM and its standard parameters, and provides a brief history of its applications. Section 4 shows M-GITM simulation results for various solar insolation and gravity wave scenarios. Section 5 summarizes the key conclusions drawn from the simulations. We discuss possible improvements that can be applied to M-GITM to better simulate the dynamics of the Martian thermosphere in Section 6.

## 2. Aphelion TPW Data Sets of Thiemann et al. (2024)

Thiemann et al. (2024) used the solar occultation measurements of temperatures and $CO_2$ densities over 100–200 km made by the EUVM instrument onboard the MAVEN spacecraft. These temperatures provide the first statistical study that identified TPW at the dawn side during aphelion conditions. The TPW data sets discussed in the work by Thiemann et al. (2024) mainly cover MY 33, 34 and 36 temperatures around aphelion near the exobase (180 km) and at 150 km. The majority of the EUVM observations cover the latitude range of 40–65°S. The data sets show a clear warming trend along the morning terminator. The temperature increases toward the polar region in these Martian years while the steepest warming trend is found in MY 34. Although the slope is different in each year, the temperature difference between the polar region and the low-latitude region is found to be generally consistent (∼40 K–60 K). In this study, the key observation that we try to reproduce with simulations is this temperature difference between the polar region and the low-latitude region along the morning terminator. A set of latitude-temperature line plots at 150 km for the morning terminator during aphelion ($L_s = 80°–100°$) is shown for each of MY 33–37 EUVM data sets (see Figure 1). The MY 34 specific data points are highlighted (in red); these will be used in data-model comparisons discussed below in Section 4.

## 3. Mars Global Ionosphere-Thermosphere Model

The Mars Global Ionosphere-Thermosphere Model (M-GITM) is a three-dimensional time-dependent non-hydrostatic GCM that solves the governing equations of momentum, energy, and mass continuity for the Martian atmosphere from the surface to ∼250 km (Bougher et al., 2015). The M-GITM code was developed from the terrestrial GITM framework (Ridley et al., 2006) by modifying it for Martian atmospheric conditions, which included implementing fundamental physical parameters, ion-neutral chemistry, key radiative processes, and solar (insolation/particle) and dynamical (wave/tidal) drivers unique to Mars (Bougher et al., 2015). The horizontal resolution of the standard model is 5° × 5° in latitude-longitude, while the vertical resolution is 2.5 km. Thus far, M-GITM has been used to interpret Mars upper atmosphere features observed using several MAVEN instruments. This includes thermospheric temperatures and key neutral densities (e.g., $CO_2$, CO, $N_2$, $O_2$, O and He) throughout Mars seasons and solar cycles 24–25, mass density distributions during Deep Dip campaigns, thermospheric winds during specific short campaigns, and 2018 global dust storm impacts on the upper atmosphere (e.g., Bougher et al. (2023); Bougher et al. (2017); Bougher, Jakosky, et al. (2015); Zurek et al. (2017); Roeten et al. (2019); Jain et al. (2020); Elrod et al. (2017, 2020); Gupta et al. (2021)).

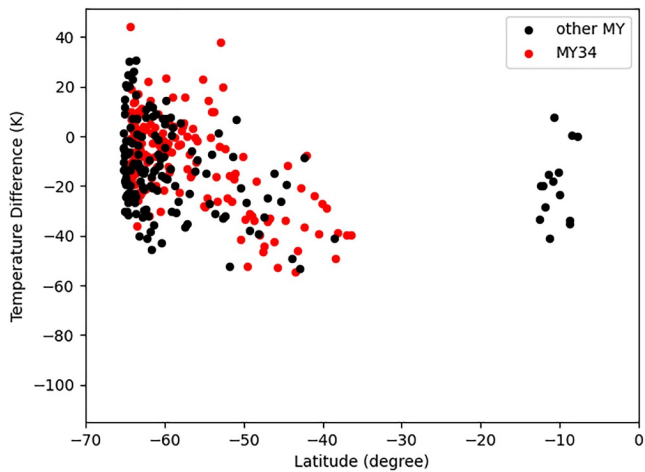

**Figure 1.** The thermosphere temperature (150 km) at the morning terminator as a function of latitude, which is observed by the EUVM instrument onboard the MAVEN during aphelion (Ls = 80–100) over several Mars Years (MY).

Required solar fluxes for driving M-GITM are taken from the MAVEN EUVM via the corresponding Flare Irradiance Spectrum Model (Mars) (FISM-M, Thiemann et al. (2017)). This empirical model provides daily averaged 1–195-nm solar fluxes throughout the MAVEN mission. The resulting M-GITM thermospheric heating, dissociation and ionization rates are computed at each model time step of 10 s (Bougher et al., 2015).

The Martian thermosphere is substantially driven by processes of lower atmospheric origin, such as dust storms and internal non-orographic gravity waves (Liu et al., 2023, 2025; Yiğit, 2023, 2024; Yiğit et al., 2021). Model inputs for time varying dust optical depths and dust mixing ratio distributions are taken from Mars Reconnaissance Orbiter Mars Climate Sounder (MCS) data sets (V6.1) (Forget et al., 1999; Kass et al., 2020; Madeleine et al., 2011; Millour et al., 2014). Due to the lack of MCS dust observations near the surface, the dust optical depth and mixing ratio are extrapolated to the surface under the well-mixed assumption (Kass et al., 2020). Specifically, two methods are employed (and model outputs compared) to capture these near aphelion MY 34 dust distributions for new M-GITM simulations. In the first method, both zonal and daily averaged values for these parameters are extracted as a function of standard MCS pressures (103 levels) and latitudes (36 elements). This 2-D grid is chosen to match the grid of latitudes (36 elements) of M-GITM, while assuming that zonal averaging is a good first option (Jain et al., 2020). In the second method, the fully 3-D dust distributions are extracted along the grid of pressures (103 levels), latitudes (36 elements) and longitudes (72 elements) matching that of M-GITM. The 3-D dust distribution is a new feature added to M-GITM. For both methods, interpolation in time between even sol intervals is conducted to provide a smooth transition between sol points for each model time step. Due to the limited number of MCS dust observations, there may be some missing values on the grid points for the 3-D dust distributions. In such cases, the missing values are filled using the average of the neighboring data points. In M-GITM, dust is treated as an aerosol with a lognormal size distribution with a mean size and variance of 2.7 μm and 0.38, respectively. The aerosol scattering properties in the visible is adopted from Ockert-Bell et al. (1997), while Mie code with the wavelength-dependent refractive index of Forget et al. (1999) is applied to the aerosol in the infrared channel. Aerosol heating rates are subsequently computed based upon these specified/empirical dust distributions.

In order to further address Mars' lower and upper atmosphere coupling, an existing whole atmosphere gravity wave parameterization (Medvedev et al., 2015; Yiğit et al., 2008) was recently incorporated, implemented, and fully tested within the M-GITM framework. It is a spectral non-linear scheme for the treatment of non-orographic gravity waves that propagate upwards to the thermosphere from their launching point at the top of the mean planetary boundary layer. It accounts for upper atmosphere processes including wave dissipation due to molecular viscosity, molecular thermal conduction, radiative damping, and nonlinear interactions between gravity waves (Roeten et al., 2022; Yiğit et al., 2008). This gravity wave scheme is now utilized when conducting specific M-GITM simulations presented in this paper. Standard gravity model parameters utilized in M-GITM are the same as those discussed in detail in the work by Roeten et al. (2022) and are commonly used in previous Martian general circulation modeling studies of gravity propagation and dissipation (Medvedev et al., 2013; Shaposhnikov et al., 2022). The two most important thermospheric impacts resulting from application of this new gravity scheme are: (a) the reduction of global winds by nearly a factor of two for all Mars seasons, and (b) the cooling of temperatures above ∼100 km at all latitudes, with the most significant cooling at high/polar latitudes.

The GW scheme has been routinely used in a variety of Earth and Mars GCMs as a standard package (Miyoshi & Yiğit, 2019; Roeten et al., 2022; Yiğit, Medvedev, & Ern, 2021; Yiğit et al., 2009).

To simulate the strong aphelion TPW during MY 34, our simulation starts from 1 November 2017 and ends on 7 December 2017. The model data for the typical variables, for example, neutrals temperature, horizontal wind velocity components, etc., are output every 2 hrs. The MCS dust profiles are obtained from the same time period. Considering that the initial conditions no longer have an impact on the simulation results after about 20 sols (Pilinski et al., 2023), we run M-GITM with a global uniform dust background (optical depth equals 0.1) 20 days before the start date to reach an equilibrium. The simulation case with the aforementioned model setup will be referred to as the standard case in the following sections.

## 4. Martian General Circulation Model Results

### 4.1. Standard Case: Simulation With Gravity Waves at Solar Minimum During Aphelion

In our modeling approach, the simulation at solar minimum during aphelion including subgrid-scale gravity waves is called the "standard simulation case." The dynamical and thermal effects of subgrid-scale gravity waves are accounted for based on the whole atmosphere gravity wave parameterization (Medvedev & Yiğit, 2012; Yiğit et al., 2008). Here we first show representative model results of the standard case for temperature at a representative time (12:00 UT on 1 December, Ls ≈ 95°). The latitude profiles of the neutral temperature at ∼150 km at the morning terminator are shown in Figure 2a. Due to actual EUVM sampling occurring slightly dayward of the morning terminator, temperature at three representative solar zenith angles are chosen for illustration: $\chi = 88°, 89°, 90°$. The southernmost points of the morning terminator profiles, located at 62.5°S latitude, are highlighted by a red box. The temperature profiles from 62.5°S to 40°S (highlighted by the blue box) can be directly compared with the observation shown in Figure 1. To illustrate the vertical structure of aphelion TPW, the vertical temperature profile along the morning terminator is presented in Figure 2b. In our standard case, we find that the temperature difference between the south polar region and the coldest low-latitude region is only around 10 K–15 K, which suggests that the MY34 aphelion TPW features cannot be reproduced by the standard case. We also perform a test run without the dust storm distribution mentioned in the previous section. The test results (not shown) are very similar to our standard case, which indicates that the local dust storms are unrelated to this aphelion TPW.

In order to investigate why the standard simulation is not able to reproduce aphelion TPW, we comprehensively analyze the dynamical heating processes in the southern polar region. Figure 2c shows the solar extreme ultraviolet (EUV) heating rate distribution in the thermosphere at 150 km. The black line illustrates the location of the morning terminator. The red box, which corresponds to the southernmost point along the near-terminator profiles in Figure 2a, highlights the area of TPW. The same annotations are applied to all other panels (from Figures 2d–2h). The total dynamical heating (Q) is contributed by 4 terms, horizontal adiabatic heating ($Q_{ha}$), horizontal hydrodynamical heating ($Q_{hh}$), vertical adiabatic heating ($Q_{va}$) and vertical hydrodynamical heating ($Q_{vh}$) (Ridley et al., 2006), which are given by:

$$Q = Q_{ha} + Q_{hh} + Q_{va} + Q_{vh}, \text{where} \tag{1}$$

$$Q_{ha} = -\left[\frac{u_\phi}{r\cos\theta}\frac{\partial\tau}{\partial\phi} + \frac{u_\theta}{r}\frac{\partial\tau}{\partial\theta}\right] \tag{2}$$

$$Q_{hh} = -\left[(\gamma-1)\tau\left(\frac{1}{r}\frac{\partial u_\theta}{\partial\theta} + \frac{1}{r\cos\theta}\frac{\partial u_\phi}{\partial\phi}\right)\right] \tag{3}$$

$$Q_{va} = -\left[u_r\frac{\partial\tau}{\partial r}\right] \tag{4}$$

$$Q_{vh} = -\left[(\gamma-1)\tau\left(\frac{2u_r}{r} + \frac{\partial u_r}{\partial r}\right)\right] \tag{5}$$

where $\tau = k_B T/\mu$ is the normalized temperature with the Boltzmann constant $k_B$, the neutral temperature $T$ and the mean molecular mass $\mu$, $r$ is the distance from the center of the planet, $\theta$ is the latitude, $\phi$ is the longitude, $u_\theta$

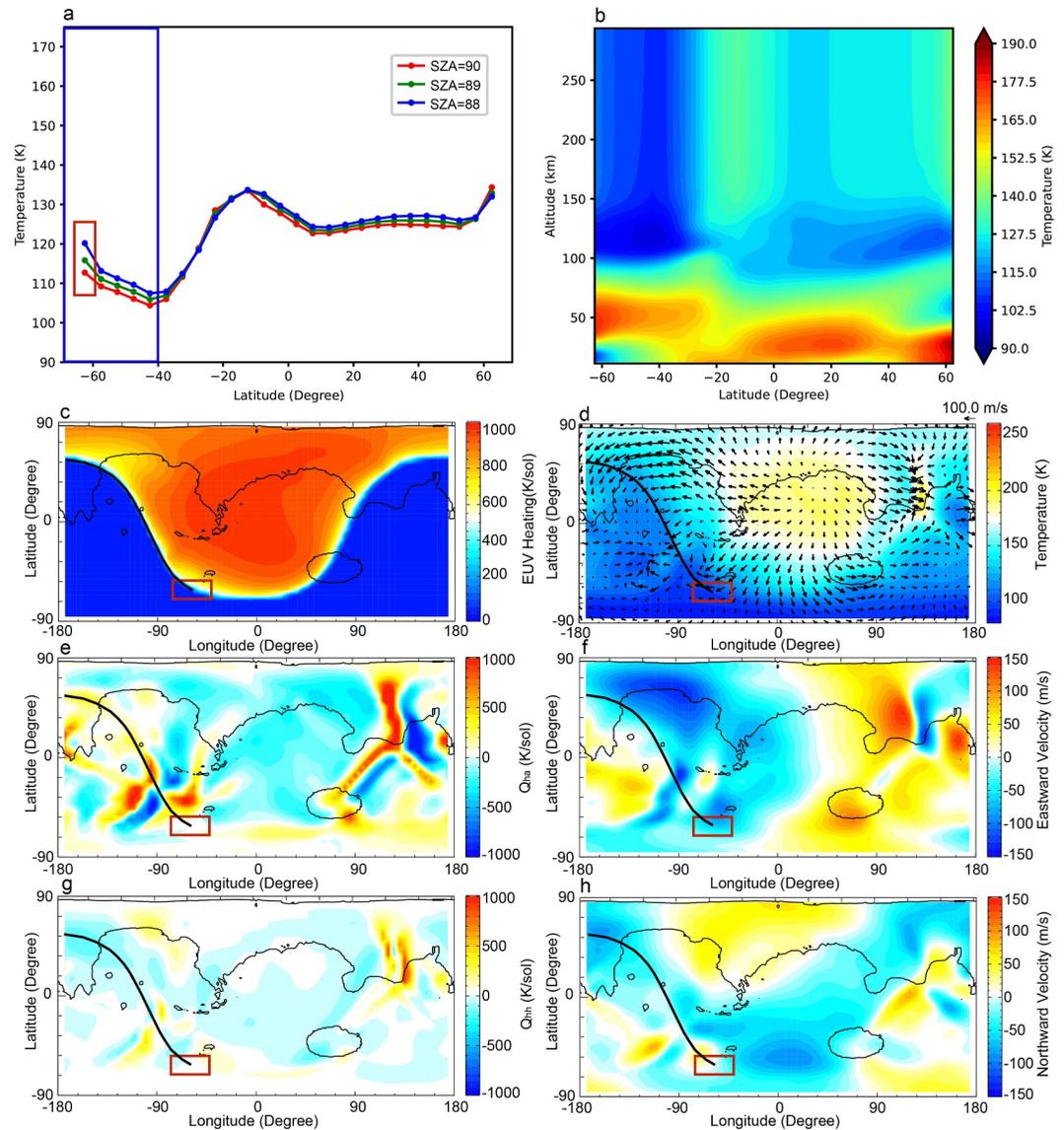

**Figure 2.** The profiles of the important atmospheric properties at the altitude of ∼150 km at 12:00 UT on 1 December (Ls ≈ 95°), extracted from the standard case simulation. (a) The thermosphere temperature profiles (150 km) near the morning terminator as functions of latitude. The solar zenith angles corresponding to each profile are shown in the legend. (b) The vertical temperature profile along the morning terminator. (c) The solar extreme ultraviolet (EUV) heating rate distribution at 150 km. (d) The horizontal temperature distribution with streamlines at 150 km. (e) The horizontal adiabatic heating rate. (f) The eastward wind velocity. (g) The horizontal hydrodynamic heating rate. (h) The northward wind velocity.

and $u_\phi$ are the meridional and zonal components of the horizontal wind velocity, respectively, and $\gamma$ is the ratio of specific heats. The advection terms (1 and 3) are proportional to the product of the wind components and the corresponding components of the temperature gradient. The divergence terms (2 and 4) are proportional to the gradients of the wind components. The value of each term is calculated by M-GITM. The 2-D temperature distribution, horizontal adiabatic heating rate, eastward wind velocity, horizontal hydrodynamic heating rate and northward wind velocity at 150 km are shown in Figures 2d–2h, respectively. Because aphelion TPW is a warming phenomenon at the morning terminator, Figures 2d–2h are presented in longitude–latitude coordinates. Since the vertical dynamical heating terms are close to 0, they are not shown in Figure 2. From Figures 2e and 2g, we can see that in the south polar region of the morning terminator, the horizontal adiabatic heating rate is 50 K–100 K/sol, while the horizontal hydrodynamic heating rate is around 0 K/sol. Combining with the temperature and velocity distributions in Figures 2d–2h, we find that the heating in this region is mainly due to the wind convergence, while the hydrodynamic advection makes little contribution to the dynamical heating. Such heating rates are not enough to generate the observed TPW. To further explore the formation mechanism of aphelion TPW, we performed the following three additional numerical experiments:

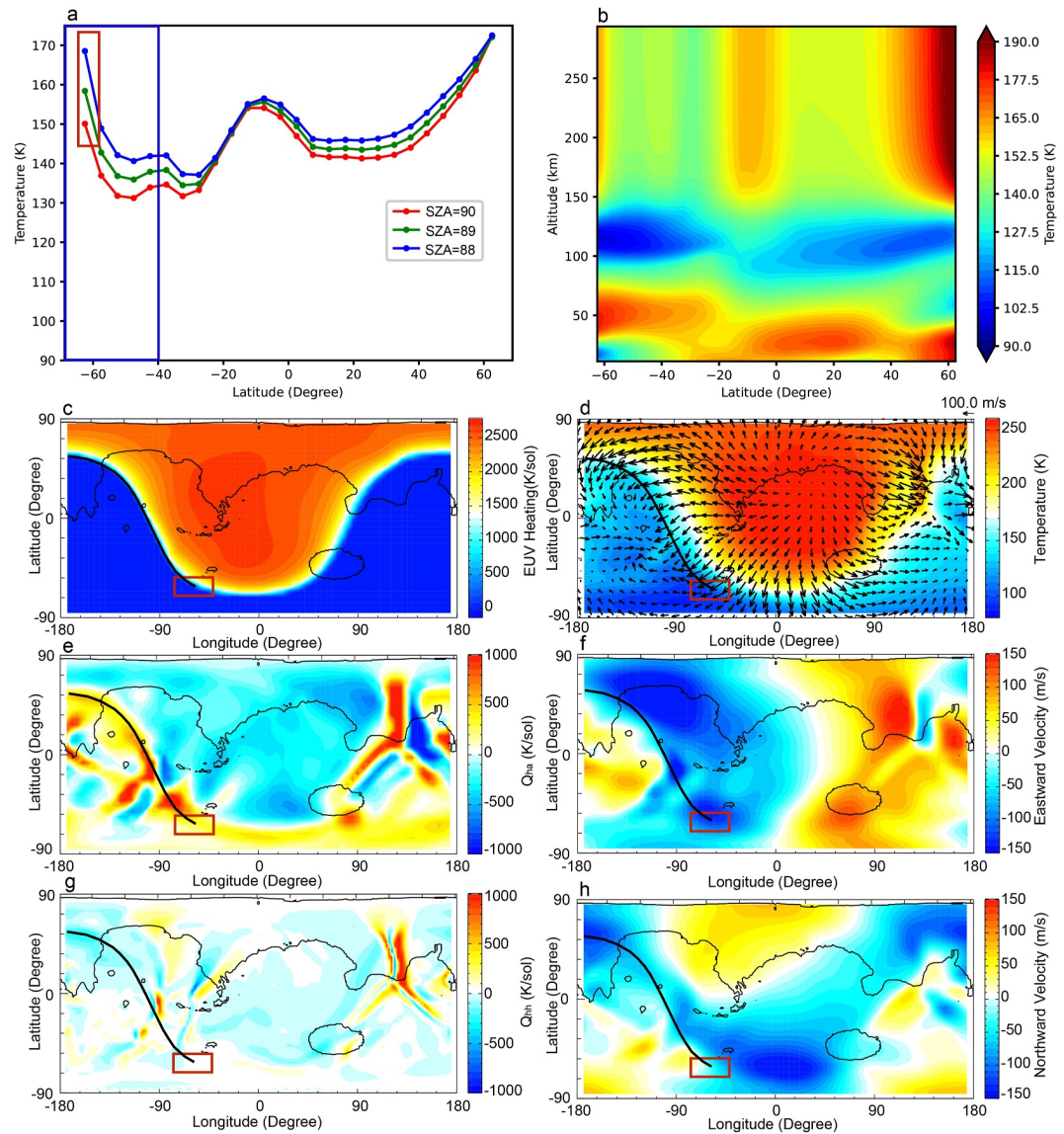

**Figure 3.** Same as Figure 2, but for the solar maximum case.

### 4.2. Study #1: Simulation With Gravity Waves at Solar Maximum During Aphelion

In our first experiment, we investigate the impact of enhanced solar insolation on aphelion TPW. We perform model simulations at solar maximum, keeping the rest of the model configurations the same as was assumed in the standard case. Model results for solar heating rate, dynamical heating terms, temperature and horizontal wind velocity are presented in Figure 3 in the same manner as in Figure 2. According to Figure 3a, we can see that with the enhanced solar insolation, the temperature difference between the thermospheric polar region and the low-latitude region at the morning terminator increases to ∼20 K–30 K. Increased solar insolation enhances horizontal circulation and temperature gradients, which in turn increase the horizontal adiabatic heating rate. Figures 3e and 3g show that the horizontal adiabatic heating rate and the horizontal hydrodynamic heating rate at the polar region are ∼300 K/sol and ∼−100 K/sol, respectively. The total (net) dynamical heating rate is ∼200 K/sol, which leads to the enhanced temperature difference across latitudes. Comparing between the solar maximum case and the standard case, we find that the global structures of the dynamical heating are similar, except that the net heating rate increases with the increasing solar insolation. The heating in the polar region is also mainly due to the wind convergence. Overall, we find that the magnitude of aphelion TPW increases with increasing solar insolation.

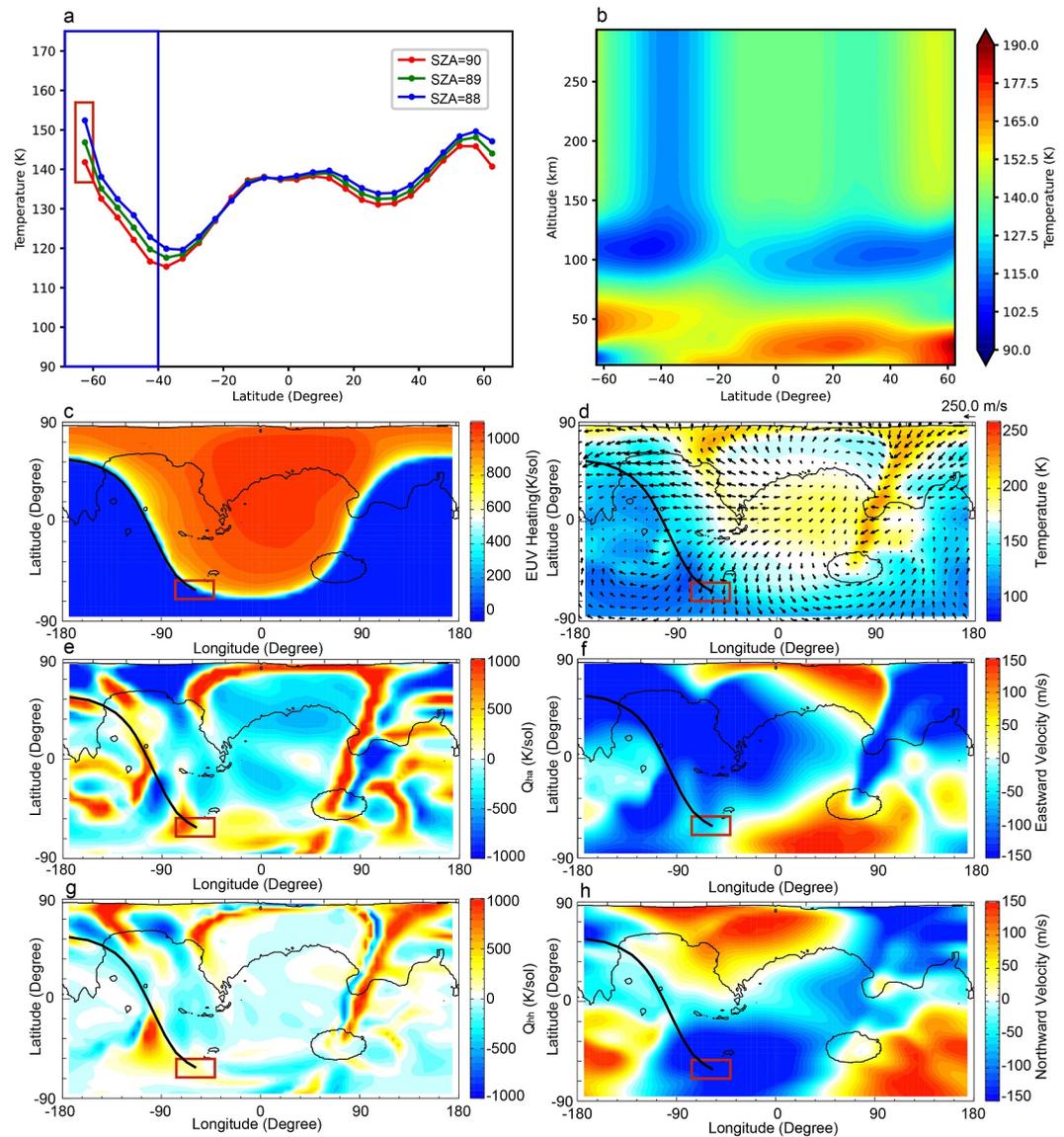

**Figure 4.** Same as Figure 2, but for the no-gravity-wave case.

### 4.3. Study #2: Simulation Without Gravity Waves at Solar Minimum During Aphelion

So far, we have included the effects of upward propagating gravity waves in the simulations. To evaluate the effect of gravity waves on the TPW, we turn off the gravity wave terms in our next sensitivity test. The rest of the model setup is the same as the standard case. The associated model outputs are shown for the same conditions as before in Figure 4. The temperature profile at the morning terminator at 150 km (Figure 4a) shows that the temperature difference across latitudes increases to ∼30 K–35 K for this new test case. Figures 4e and 4g show that the horizontal adiabatic heating rate and the horizontal hydrodynamic heating rate at the polar region are ∼300 K/sol and ∼150 K/sol, respectively. The positive horizontal hydrodynamic heating rate suggests that the wind vector is in the same direction as the horizontal temperature gradient and makes considerable contribution to aphelion TPW. The heating rate distributions and the horizontal wind maps (Figures 4e–4h) of this test case become very different from those of the standard case. Note that this hypothetical scenario of "no gravity waves" is an unrealistic case that serves solely as a sensitivity experiment. In the absence of gravity waves, middle atmospheric jets extend all the way to the thermosphere without reversing their directions, for example, as can be seen in Medvedev et al. (2013, Figure 9) and Roeten et al. (2022, Figure 2). Removing gravity waves from the model setup artificially leads to unrealistically strong winds, which can drive excessively large dynamical heating as seen by comparison of Figures 2 and 4. The enhanced dynamical heating could then be responsible for the greater latitudinal temperature difference. In this case, the identified simulation features appear to be an example of missing physics in the model leading to compensation by, or overestimation of, other physical processes that are included. In summary, this experiment shows that the gravity wave parameterization used in M-GITM tends to suppress the magnitude of aphelion TPW.

### 4.4. Study #3: Simulation With High Resolution M-GITM

In the standard case, M-GITM is set up to run with a 5° × 5° latitude-longitude grid. This coarse grid resolution may not sufficiently resolve smaller scale processes. In order to study whether variations in the resolved scale processes can influence the degree of aphelion TPW, we run M-GITM with a higher resolution of 2.5° × 2.5° latitude-longitude grid. The model setup stays the same as in the standard case except for the increased horizontal resolution. Figure 5 presents the high-resolution model results, with panels corresponding to those shown in previous figures. The temperature profile near the morning terminator (Figure 5a) shows that the temperature difference between the polar region and ∼40°S remains the same as the standard case (∼10 K–15 K). From Figures 5e and 5g, we can see that in the south polar region of the morning terminator, the horizontal adiabatic and hydrodynamic heating rates are the same as the heating rates in the standard case. However, comparing the high-resolution dynamical heating rate distributions with those of the standard case, Figures 5d and 5f show that the dynamical heating in the low-latitude region looks different in the high resolution case. The temperature profiles (Figures 5a and 5b) also show that the temperature difference between 40°S and the equator is higher than that in the standard case. Such a difference suggests that some smaller scale processes in the Martian thermosphere may not be well reproduced by a 5° × 5° latitude-longitude grid. Overall, the model horizontal resolution does not considerably impact the simulation of aphelion TPW.

## 5. Conclusions

Aphelion TPW, first identified observationally by Thiemann et al. (2024), is a spectacular manifestation of thermodynamical coupling in the Martian thermosphere. Exploring the formation mechanisms of aphelion TPW will improve our understanding of the structure of the Martian thermosphere. In this study, we pursue a methodology to study aphelion TPW using the Mars Global Ionosphere Thermosphere (M-GITM) global circulation model. We performed a suite of systematic global-scale model simulations, interactively coupled to the whole atmosphere gravity wave parameterization of Yiğit et al. (2008), and analyzed the sensitivity of aphelion TPW to various forcing mechanisms. In our standard case, which is based on a simulation performed at solar minimum with gravity waves, we find that the temperature difference between the polar region and the low-latitude region at the morning terminator is only ∼10 K–15 K, which is significantly lower than the MAVEN EUVM observations (Thiemann et al., 2024). Several test cases are performed to explore the influence of dust, solar insolation, subgrid-scale gravity waves and horizontal resolution on the latitudinal temperature difference. Based on the simulation results, we draw the following conclusions:

1. The degree of TPW in the aphelion winter hemisphere polar region is underestimated with the baseline M-GITM setup. The computed 40–65°S latitude temperature difference at 150 km is ∼20 K–30 K lower than that measured by the EUVM solar occultation technique (Thiemann et al., 2024). Such difference is likely due to the underestimation of dynamical heating in the polar region in M-GITM.
2. Regardless of which set of MY 34 aphelion dust distributions (2-D, 3-D and no dust) is used, the M-GITM model generates very similar temperature distributions near the morning terminator. Hence, our simulations imply that dust distribution and activity are unrelated to TPW for the aphelion case.
3. By increasing the solar insolation, we find that the temperature difference will increase owing to faster horizontal winds and modified wind convergence. The horizontal adiabatic heating at the polar region is considerably enhanced by the stronger solar insolation (from ∼50 K–100 K/sol to ∼300 K/sol), while the horizontal hydrodynamic heating is decreased (from ∼0 K/sol to ∼−100 K/sol). The overall dynamical heating structure in the Martian thermosphere does not change much in the solar maximum case. Nevertheless, horizontal wind magnitudes indeed regulate the magnitude of aphelion TPW heating and the resulting thermospheric temperatures at 150 km.

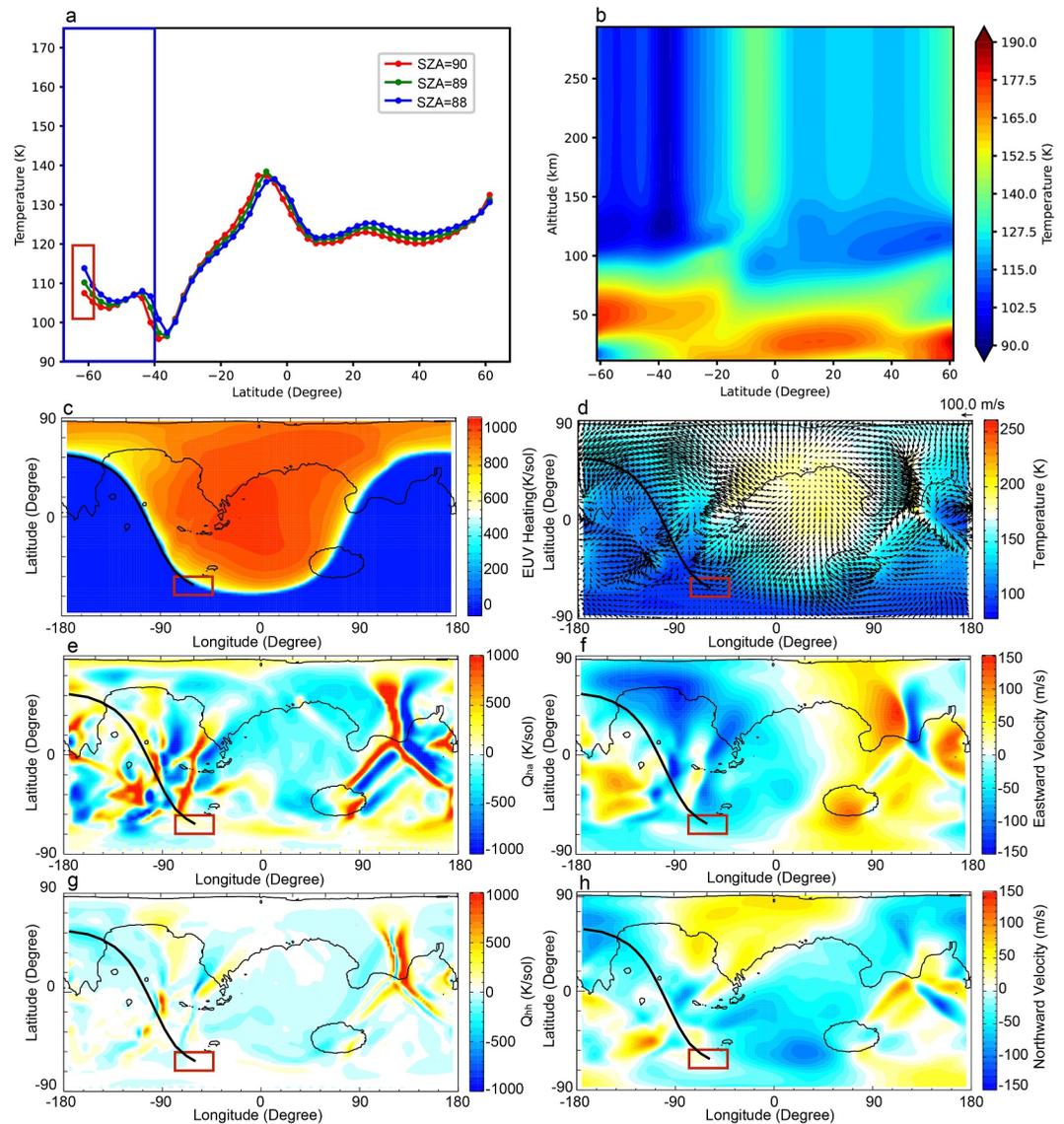

**Figure 5.** Same as Figure 2, but for the high-resolution case.

4. Gravity waves have considerable impact on aphelion TPW. By turning the gravity waves scheme off in the test run, we find that the temperature difference between the polar region and the low-latitude region increases with the enhanced horizontal hydrodynamic heating. This suggests that the dynamical heating compensates for the missing gravity waves when they are artificially removed from the model. Gravity waves generally slow down thermospheric winds, suppressing the dynamical heating especially in the polar regions.
5. The spatial resolution of M-GITM has little impact on the polar warming at 150 km, while the fine thermosphere structure of the low-latitude region may be sensitive to the spatial resolution.

# 6. Discussions

The simulation results of the standard case and the test cases suggest that the modeling study on aphelion TPW can be improved in the following aspects:

1. The use of the whole atmosphere gravity wave scheme by (Yiğit et al., 2008) as implemented by Roeten et al. (2022) in M-GITM demonstrated that the deposition of gravity wave momentum and energy in the Mars upper atmosphere considerably impacts the winds and temperature in the thermosphere. Terrestrial GCM studies utilizing the whole atmosphere parameterization showed that the implementation of a latitudinally varying gravity wave source spectrum based on global observations of gravity waves can improve the simulations of winds and temperature (Yiğit, Medvedev, & Ern, 2021). Such studies highlight the complexity of gravity wave source spectra in planetary atmospheres and a similar approach may be adopted on Mars. Namely, a recent Martian general circulation modeling study by Kling et al. (2025) prescribed a latitude dependency on the amplitude of the source of non-orographic gravity waves, with an exponential decrease polewards of 45°N/S latitude. However, there is currently a lack of knowledge of the latitude distribution of the gravity wave source spectrum associated with the dominant wavelengths of gravity waves used in MGCMs. In principle, gravity waves with different wavelengths can have a different spectral distribution. Dedicated observational campaigns are needed in the future to better characterize the gravity waves scales that are included in MGCMs. In addition, the modification of the gravity wave launch height with season and diurnal structure is important to address (e.g., Liu et al. (2023, 2025)). In short, an improved representation of gravity wave processes is needed to better capture Martian thermosphere dynamics in global 3-D models.

2. In the current model, the Martian topography is not considered, which leads to the neglect of the orographic gravity waves. Orographic gravity waves, owing to their relatively small intrinsic phase speeds and short vertical wavelengths, have small penetration depth in the atmosphere, saturate and/or are critically filtered in the Martian lower and middle atmosphere, where they can contribute to the overall momentum budget (Forget et al., 1999; Medvedev et al., 2011). Due to their unfavorable upward propagation conditions and their tendency to become unstable at lower altitudes, the thermospheric effects of orographic gravity waves are negligible compared to the non-orographic ones. Nevertheless, the joint application of both orographic and non-orographic schemes within M-GITM is warranted, since lower atmosphere horizontal wind modification by orographic waves can potentially modulate the upward propagation of non-orographic waves on their path to the thermosphere (Kling et al., 2025) via changes in the mean winds.

3. In global circulation models, including M-GITM, polar filters are applied to mitigate numerical instabilities near the poles. These instabilities arise primarily due to the convergence of meridians in spherical coordinate systems, where the grid spacing becomes extremely small in the east-west direction as one approaches the poles. The polar filter acts as a numerical damper, typically removing or reducing high-wavenumber components near the poles. The high-frequency components in the high-latitude region may play an important role in the formation of aphelion TPW. Recently, the NASA Ames Mars Global Climate Model adopted the cubed-sphere dynamical core from Geophysical Fluid Dynamics Laboratory's FV3 model (Kling et al., 2025), offering a nearly uniform global grid that eliminates the need for polar filters. A simple approach of implementing cubed-sphere geometry was proposed by Chen and Li (2024). A potential future direction involves upgrading the M-GITM model to a cubed-sphere grid and investigating whether aphelion TPW is simulated better in a GCM that does not require polar filtering.

Thiemann et al. (2024), which first identified aphelion TPW, tried to study this phenomenon with the Mars Climate Database, but the simulated temperature difference is also considerably lower than the value observed. Since the simulation results from M-GITM do not well reproduce aphelion TPW either, current Mars global circulation models may be missing some important physical process in the Martian atmosphere. Our work suggests that the ability to accurately reproduce this phenomenon may be considered a key indicator of a future Mars model's performance and reliability.

## Conflict of Interest

The authors declare no conflicts of interest relevant to this study.

## Availability Statement

MAVEN EUVM data are publicly available at the NASA Planetary Data System available at Eparvier et al. (2023). The M-GITM code used to generate our model results is open-source and publicly available online (https://github.com/dpawlows/MGITM, Bougher, Pawlowski, et al. (2015)). The model outputs used to generate the figures in this study are available at https://doi.org/10.7302/5vj1-qh36.

## Acknowledgments

This work is supported by the Science and Technology Development Fund (FDCT) of Macau (Grants: 0095/2025/ITP2 and 002/2024/SKL). CL is supported by the Heising-Simons Foundation.

## References

Bell, J. M., Bougher, S. W., & Murphy, J. R. (2007). Vertical dust mixing and the interannual variations in the mars thermosphere. *Journal of Geophysical Research*(E12), 112. 1991–2012.

Bougher, S. W., Bell, J. M., Murphy, J. R., Lopez-Valverde, M. A., & Withers, P. G. (2006). Polar warming in the mars thermosphere: Seasonal variations owing to changing insolation and dust distributions. *Geophysical Research Letters*, *33*(2). https://doi.org/10.1029/2005gl024059

Bougher, S. W., Benna, M., Elrod, M., Roeten, K. J., & Thiemann, E. (2023). MAVEN/NGIMS dayside exospheric temperatures over solar cycle and seasons: Role of dayside thermal balances in regulating temperatures. *Journal of Geophysical Research: Planets*, *128*(1), e2022JE007475. https://doi.org/10.1029/2022JE007475

Bougher, S. W., Jakosky, B. M., Halekas, J., Grebowsky, J., Luhmann, J. G., et al. (2015). Early MAVEN dip deep campaign reveals thermosphere and ionosphere variability. *Science*, *350*, 1–7. https://doi.org/10.1126/science.aad0459

Bougher, S. W., Pawlowski, D., Bell, J. M., Nelli, S., McDunn, T., Murphy, et al. (2015). Mars Global Ionosphere-Thermosphere Model: Solar cycle, seasonal, and diurnal variations of the Mars upper atmosphere. *Journal of Geophysical Research*, *120*(2), 311–342. https://doi.org/10.1002/2014JE004715

Bougher, S. W., Roeten, K. J., Olsen, K., Mahaffy, P. R., Benna, M., Elrod, M., et al. (2017). The structure and variability of Mars dayside thermosphere from MAVEN NGIMS and IUVS measurements: Seasonal and solar activity trends in scale heights and temperatures. *Journal of Geophysical Research: Space Physics*, *122*(1), 1296–1313. https://doi.org/10.1002/2016JA023454

Brecht, A. S., Bougher, S. W., Gérard, J. C., Parkinson, C. D., And, S. R., & Foster, B. (2011). Understanding the variability of nightside temperatures, no uv and o2 ir nightglow emissions in the venus upper atmosphere. *Journal of Geophysical Research*.

Chen, S., & Li, C. (2024). Exocubed: A Riemann-solver-based cubed-sphere dynamic core for planetary atmospheres. *The Astrophysical Journal*, *966*(1), 123. https://doi.org/10.3847/1538-4357/ad33b9

Crowley, G., Schoendorf, J., & Marcos, F. (1995). *Satellite observations of neutral density cells in the lower thermosphere at high latitudes* (Vol. 87, pp. 339–347). agu geophysical monograph on the upper mesosphere and lower thermosphere.

Elrod, M. K., Bougher, S., Bell, J., Mahaffy, P. R., Benna, M., Stone, S., et al. (2017). He bulge revealed: He and CO2 diurnal and seasonal variations in the upper atmosphere of Mars as detected by MAVEN NGIMS. *Journal of Geophysical Research: Space Physics*, *122*(2), 2564–2573. https://doi.org/10.1002/2016JA023482

Elrod, M. K., Bougher, S. W., Roeten, K., Sharrar, R., & Murphy, J. (2020). Structural and compositional changes in the upper atmosphere related to the PEDE-2018 dust event on Mars as observed by MAVEN NGIMS. *Geophysical Research Letters*, *47*(4), e2019GL084378. https://doi.org/10.1029/2019GL084378

Eparvier, F. G., Thiemann, E. M. B., Chamberlin, P. C., Templeman, B., & Borelli, R. (2023). The maven euvm level 2 data (version 14) [Dataset]. *NASA PDS*. Retrieved from https://pds-ppi.igpp.ucla.edu/mission/MAVEN/MAVEN/EUV

Forbes, J. M., & Moudden, Y. (2009). Solar terminator wave in a mars general circulation model. *Geophysical Research Letters*, *36*(17), 1397–1413. https://doi.org/10.1029/2009gl039528

Forget, F., Hourdin, F., Fournier, R., Hourdin, C., Talagrand, O., Collins, M., et al. (1999). Improved general circulation models of the martian atmosphere from the surface to above 80 km. *Journal of Geophysical Research*, *104*(E10), 24155–24175. https://doi.org/10.1029/1999JE001025

González-Galindo, F., Forget, F., López-Valverde, M. A., & Angelats i Coll, M. (2009). A ground-to-exosphere Martian general circulation model: 2. Atmosphere during solstice conditions–thermospheric polar warming. *Journal of Geophysical Research (Planets)*, *114*(E13), E08004. https://doi.org/10.1029/2008JE003277

Gupta, N., Rao, N. V., Bougher, S. W., & Elrod, M. K. (2021). Latitudinal and seasonal asymmetries of the helium bulge in the Martian upper atmosphere. *Journal of Geophysical Research: Planets*, *126*(10), e2021JE006976. https://doi.org/10.1029/2021JE006976

Heavens, N. G., Pankine, A., Battalio, J. M., Wright, C., Kass, D. M., Kleinböhl, A., et al. (2022). Mars climate sounder observations of gravity-wave activity throughout Mars's lower atmosphere. *The Planetary Science Journal*, *3*(3), 57. https://doi.org/10.3847/PSJ/ac51ce

Jain, S. K., Bougher, S. W., Deighan, J., Schneider, N. M., González Galindo, F., Stewart, A. I. F., et al. (2020). Martian thermospheric warming associated with the Planet Encircling Dust Event of 2018. *Geophysical Research Letters*, *47*(3), e2019GL085302. https://doi.org/10.1029/2019GL085302

Jesch, D., Medvedev, A. S., Castellini, F., Yiğit, E., & Hartogh, P. (2019). Density fluctuations in the lower thermosphere of Mars retrieved from the ExoMars Trace Gas Orbiter (TGO) aerobraking. *Atmosphere*, *10*(10), 620. https://doi.org/10.3390/atmos10100620

Kass, D., Schofield, J., Kleinböhl, A., McCleese, D., Heavens, N., Shirley, J., & Steele, L. (2020). Mars climate sounder observation of mars' 2018 global dust storm. *Geophysical Research Letters*, *47*(23), e2019GL083931. https://doi.org/10.1029/2019gl083931

Kling, A., Wilson, R. J., Kahre, M., Brecht, A., & Murphy, J. (2025). Impact of grid resolution on wave-mean flow interactions with high resolution Mars global climate model simulations. *Geophysical Research Letters*, *52*(2), e2024GL112297. https://doi.org/10.1029/2024GL112297

Liu, J., Millour, E., Forget, F., Gilli, G., Lott, F., Bardet, D., et al. (2023). A surface to exosphere non-orographic gravity wave parameterization for the mars planetary climate model. *Journal of Geophysical Research: Planets*, *128*(7), e2023JE007769. https://doi.org/10.1029/2023je007769

Liu, J., Millour, E., Forget, F., Gilli, G., Lott, F., Bardet, D., & Galindo, F. G. (2025). Diurnal cycle of non-orographic gravity waves' source altitudes and its impacts: Tests with mars planetary climate model. *Journal of Geophysical Research: Planets*, *130*(7), e2024JE008880. https://doi.org/10.1029/2024je008880

Madeleine, J.-B., Forget, F., Millour, E., Montabone, L., & Wolff, M. (2011). Revisiting the radiative impact of dust on mars using the lmd global climate model. *Journal of Geophysical Research*, *116*(E11), E11010. https://doi.org/10.1029/2011je003855

Medvedev, A. S., González-Galindo, F., Yiğit, E., Feofilov, A. G., Forget, F., & Hartogh, P. (2015). Cooling of the Martian thermosphere by $CO_2$ radiation and gravity waves: An intercomparison study with two general circulation models. *Journal of Geophysical Research: Planets*, *120*(5), 913–927. https://doi.org/10.1002/2015JE004802

Medvedev, A. S., & Yiğit, E. (2012). Thermal effects of internal gravity waves in the Martian upper atmosphere. *Geophysical Research Letters*, *39*(5). https://doi.org/10.1029/2012GL050852

Medvedev, A. S., Yiğit, E., & Hartogh, P. (2011). Estimates of gravity wave drag on Mars: Indication of a possible lower thermosphere wind reversal. *Icarus*, *211*(1), 909–912. https://doi.org/10.1016/j.icarus.2010.10.013

Medvedev, A. S., Yiğit, E., Kuroda, T., & Hartogh, P. (2013). General circulation modeling of the martian upper atmosphere during global dust storms. *Journal of Geophysical Research: Planets*, *118*(10), 1–13. https://doi.org/10.1002/2013JE004429

Millour, E., Forget, F., & Lewis, S. (2014). Mars climate database v. 5.1. user manual. *ESTEC Contract*, *11369*, 95.

Miyoshi, Y., & Yiğit, E. (2019). Impact of gravity wave drag on the thermospheric circulation: Implementation of a nonlinear gravity wave parameterization in a whole-atmosphere model. *Annales Geophysicae*, *37*(5), 955–969. https://doi.org/10.5194/angeo-37-955-2019

Ockert-Bell, M. E., Bell III., J. F., Pollack, J. B., McKay, C. P., & Forget, F. (1997). Absorption and scattering properties of the martian dust in the solar wavelengths. *Journal of Geophysical Research*, *102*(E4), 9039–9050. https://doi.org/10.1029/96je03991

Pilinski, M., Bougher, S. W., Greer, K., Thiemann, E., Andersson, L., Benna, M., & Elrod, M. (2018). First evidence of persistent night-time temperature structures in the neutral thermosphere of Mars. *Geophysical Research Letters*, *45*(17), 8819–8825. https://doi.org/10.1029/2018GL078761

Pilinski, M. D., Roeten, K. J., Bougher, S. W., & Benna, M. (2023). Dynamical heating in the martian thermosphere. *Journal of Geophysical Research: Planets*, *128*(6), e2022JE007670. https://doi.org/10.1029/2022je007670

Ridley, A., Deng, Y., & Tòth, G. (2006). The global ionosphere-thermosphere model. *Journal of Atmospheric and Solar-Terrestrial Physics*, *68*(8), 839–864. https://doi.org/10.1016/j.jastp.2006.01.008

Roeten, K. J., Bougher, S. W., Benna, M., Elrod, M., Medvedev, A., & Yiğit, E. (2022). Impacts of gravity waves in the Martian thermosphere using M-GITM coupled with a whole atmosphere gravity wave scheme. *Journal of Geophysical Research: Planets*, *127*(12), e2022JE007477. https://doi.org/10.1029/2022JE007477

Roeten, K. J., Bougher, S. W., Benna, M., Mahaffy, P. R., Lee, Y., Pawlowski, D., et al. (2019). MAVEN/NGIMS thermospheric neutral wind observations: Interpretation using the M-GITM general circulation model. *Journal of Geophysical Research: Planets*, *124*(12), 3283–3303. https://doi.org/10.1029/2019JE005957

Schoendorf, J., Crowley, G., & Roble, R. (1996). Neutral density cells in the high latitude thermosphere - 2. Mechanisms. *Journal of Atmospheric and Terrestrial Physics*, *58*(15), 1769–1781. https://doi.org/10.1016/0021-9169(95)00166-2

Shaposhnikov, D. S., Medvedev, A. S., Rodin, A. V., Yiğit, E., & Hartogh, P. (2022). Martian dust storms and gravity waves: Disentangling water transport to the upper atmosphere. *Journal of Geophysical Research: Planets*, *127*(1), e2021JE007102. https://doi.org/10.1029/2021JE007102

Starichenko, E. D., Medvedev, A. S., Belyaev, D. A., Yiğit, E., Fedorova, A. A., Korablev, O. I., et al. (2024). Climatology of gravity wave activity based on two martian years from ACS/TGO observations. *Astronomy & Astrophysics*, *683*, A206. https://doi.org/10.1051/0004-6361/202348685

Thiemann, E. M. B., Chamberlin, P. C., Eparvier, F. G., Templeman, B., Woods, T. N., Bougher, S. W., & Jakosky, B. M. (2017). The MAVEN EUVM model of solar spectral irradiance variability at Mars: Algorithms and results. *Journal of Geophysical Research: Space Physics*, *122*(3), 2748–2767. https://doi.org/10.1002/2016JA023512

Thiemann, E. M. B., Trompet, L., Bougher, S. W., Yiğit, E., Gasperini, F., Montabone, L., et al. (2024). The climatology of mars thermospheric polar warming at aphelion. *Geophysical Research Letters*, *51*(5), e2023GL107140. https://doi.org/10.1029/2023gl107140

Yiğit, E. (2023). Coupling and interactions across the Martian whole atmosphere system. *Nature Geoscience*, *16*(2), 123–132. https://doi.org/10.1038/s41561-022-01118-7

Yiğit, E. (2024). Exploring Mars's harsh atmosphere. *Physics Today*, *77*(7), 42–50. https://doi.org/10.1063/pt.sjcv.azbb

Yiğit, E., Aylward, A. D., & Medvedev, A. S. (2008). Parameterization of the effects of vertically propagating gravity waves for thermosphere general circulation models. *Sensitivity study*, *113*, D19106. https://doi.org/10.1029/2008JD010135

Yiğit, E., Medvedev, A. S., & Hartogh, P. (2021). Variations of the martian Thermospheric gravity-wave activity during the recent solar minimum as observed by MAVEN. *The Astrophysical Journal*, *920*(2), 69. https://doi.org/10.3847/1538-4357/ac15fc

Yiğit, E., & Medvedev, A. S. (2019). Obscure waves in planetary atmospheres. *Physics Today*, *72*(6), 40–46. https://doi.org/10.1063/PT.3.4226

Yiğit, E., Medvedev, A. S., Aylward, A. D., Hartogh, P., & Harris, M. J. (2009). Modeling the effects of gravity wave momentum deposition on the general circulation above the turbopause. *Journal of Geophysical Research*, *114*(D7), D07101. https://doi.org/10.1029/2008JD011132

Yiğit, E., Medvedev, A. S., Benna, M., & Jakosky, B. M. (2021). Dust storm-enhanced gravity wave activity in the martian thermosphere observed by maven and implication for atmospheric escape. *Geophysical Research Letters*, *48*(5), e2020GL092095. https://doi.org/10.1029/2020GL092095

Yiğit, E., Medvedev, A. S., & Ern, M. (2021). Effects of latitude-dependent gravity wave source variations on the middle and upper atmosphere. *Frontiers in Astronomy and Space Sciences*, *7*, 614018. https://doi.org/10.3389/fspas.2020.614018

Zurek, R. W., Tolson, R. A., Bougher, S. W., Lugo, R. A., Baird, D. T., Bell, J. M., & Jakosky, B. M. (2017). Mars thermosphere as seen in MAVEN Accelerometer data. *Journal of Geophysical Research: Space Physics*, *122*(3), 3798–3814. https://doi.org/10.1002/2016JA023641